\documentclass[11pt]{article}

\usepackage[utf8]{inputenc}      % omit under lualatex/xelatex

\usepackage[T1]{fontenc}

\usepackage{amsthm}
\usepackage{newtxtext,newtxmath}

\usepackage{microtype}

\usepackage[a4paper, margin=1.2in]{geometry}
\usepackage[round,authoryear,sort&compress]{natbib}

\usepackage{xcolor}
\definecolor{linknavy}{HTML}{1A3D6D}
\usepackage{hyperref}
\hypersetup{
  colorlinks=true,               % swap for `hidelinks` if you want black links
  linkcolor=linknavy,
  citecolor=linknavy,
  urlcolor=linknavy,
  breaklinks=true,
  pdftitle={A Note on the Strategic Confinement Problem},
  pdfauthor={Christian Schroeder de Witt},
  pdfsubject={Confinement, multi-agent security, correlated equilibrium}
}
\usepackage{cleveref}

\theoremstyle{definition}

\crefname{observation}{Observation}{Observations}
\Crefname{observation}{Observation}{Observations}

\title{A Note on the Strategic Confinement Problem}

\author{Christian Schroeder de Witt\thanks{%
  University of Oxford. \texttt{contact@wittlab.ai}}}

\date{\today}            
\begin{document}
\maketitle

\begin{abstract}
Lampson's confinement problem asks how to prevent a program that processes confidential information from leaking it to a third party. We introduce the strategic confinement problem, which arises when the communicating parties are strategic agents with shared coordination resources. In this setting, residual communication capacity can be concentrated on low-entropy, high-impact predicates of the confidential data. Consequently, bounds on information leakage need not induce corresponding bounds on worst-case harm: a channel with negligible capacity may still suffice to select damaging outcomes. We argue that systems of learnt strategic agents naturally instantiate this problem because they do not admit complete behavioural specifications, their learnt conventions generally cannot be predicted or reproduced by an external observer, and sufficiently capable agents can construct covert communication schemes that are difficult to detect or eliminate. Our contribution is therefore not a new theory of communication, but a reinterpretation of confinement in the presence of strategic agents. Classical confinement bounds what information may flow; strategic confinement highlights that this need not bound what strategic agents can jointly achieve.
\end{abstract}

% =====================================================================
% MAIN NOTE  (unsectioned lead — faithful to Lampson's note form)
% Arc: Lampson's confinement problem -> lift to learnt strategic agents
%      -> central claim (Observation: strategic confinement gap)
%      -> why learnt-agent systems admit it -> bound harm directly.
%
% If you want navigation, uncomment light section headers, e.g.:
% \section{Introduction}
% \section{Lifting Lampson's setting to strategic agents}
% \section{Strategic confinement in systems of learnt agents}
% =====================================================================

% [main argument]
%
% At the central claim, use:
% \begin{observation}[Strategic confinement gap]\label{obs:gap}
%   ...
% \end{observation}

\section{The Confinement Problem}

In 1973, Lampson published a brief note on the confinement problem~\citep{lampson-confinement}. On the basis of a simple setting involving multiple communicating programs in a shared multi-user environment, he examines under what conditions it can be guaranteed that confidential data is not leaked. Despite its brevity, the note has had a profound influence on computer security as it exposed a fundamental limitation of access control: once allowed to observe confidential data, ensuring that this data is not leaked may require reasoning over covert channels rather than merely sanctioned interfaces. In this paper, we lift Lampson's setting from specified programs to learnt strategic agents, and ask what follows for the design of secure systems of artificially intelligent agents.

Lampson's setting consists of a customer program, a service program, and a service owner. To maintain system utility, the customer needs to share confidential customer data with the service program. The customer then needs to settle a bill for the services received. A copy of the bill needs to be shared with the service owner. The service owner may not receive confidential customer data. Lampson's confinement problem then asks whether it can be guaranteed that no confidential customer data is shared with the service owner while maintaining the system's full utility.

Lampson's key insight is that the confinement problem requires reasoning over covert channels. If the service program generates the bill, then it might hide confidential customer information in the exact bill amount. The service owner might then recover that information once receiving a copy of the bill. To guarantee confinement, every possible channel needs to be identified and removed. Sometimes, this is possible while maintaining the system's utility: the bill could be generated by the customer and either accepted or rejected by the service owner, as in modern procurement systems. However, Lampson suspects that guaranteeing confinement in even moderately complicated~\footnote{Lampson uses the term ``systems of [...] moderate complexity''. We here refrain from using the term “complex” in favour of  ``complicated'' to avoid making a complexity-theoretic statement or alluding to the complex-systems literature.} systems may be routinely impossible. Lampson suggests that if a covert channel cannot be removed, then its capacity should be reduced as far as utility permits.

\section{The Strategic Confinement Problem}

We will now attempt to lift Lampson's confinement problem from specified programs to learnt strategic agents. Our central observation is that information-theoretic confinement guarantees only imply harm-theoretic guarantees under additional assumptions relating information leakage to damage. Here a harm-theoretic guarantee is a bound on the worst-case harm to the customer, as against an information-theoretic guarantee, which bounds the bits the owner can recover. Strategic agents can invalidate these assumptions because equilibrium-selection resources allow tiny signals to unlock highly damaging coordinated actions. 

Note that we do not argue that classical confinement guarantees are incorrect. Rather, we argue that when communicating parties are strategic, information-theoretic confinement guarantees need not induce corresponding harm-theoretic guarantees.

As a first step, assume the bill creation routine is contained within a complicated proprietary legacy program that makes it economically prohibitive to rewrite or legally cumbersome to share. Let us assume that this prevents the bill subroutine from being certified, or shared with, or executed by, the customer. Further assume that the customer does not have the skill or cognitive resources to check that the bill is exactly correct but merely that it is roughly plausible. In this case, clearly the customer loses the ability to generate the bill herself. She also cannot mask the service process while simultaneously guaranteeing the utility of the system.

As a consequence, in Lampson's classical confinement setting the customer can no longer guarantee confinement; the worst-case information the service owner can recover about her confidential data is upper-bounded by the estimated capacity of the covert channel formed between the service process and its owner, multiplied by the number of messages exchanged through it. This leakage bound is the only confinement guarantee she retains. 

We now advance further toward our central claim:  \emph{information-theoretic confinement guarantees do not generally induce
  harm-theoretic confinement guarantees for strategic agents, and the implications of this distinction for confinement have been understudied\footnote{See the Appendix for a placement in the prior literature.}.}
% \begin{observation}[Strategic confinement gap]\label{obs:gap}
 
% \end{observation}

To this end, we further assume that the service and owner programs are \textit{strategic} programs, meaning they are endowed with reasoning capabilities, as follows. In particular, we assume that service and owner programs share a non-zero amount of common knowledge about the natural world, and that this naturally gives rise to a certain amount of joint coordination resources in the form of preferred common knowledge labels, or Schelling points. Given an input, such as the confidential customer data, and its coordination resources, the service can then reason over the channel output, namely the bill. Likewise, the owner can reason over their channel input, namely the bill, together with the joint coordination resources.

Once even a small amount of reliable signalling remains, the security loss is governed not only by channel capacity but also by the agents' shared ability to select high-impact propositions for transmission. If the service program is not assumed strategic, we have no reason to expect it to preferentially leak the parts of the customer data that cause the most harm, such as an aggregated credit score. If instead the service and owner programs are strategic and share suitable coordination resources, then whenever some low-entropy, high-impact function of the confidential data exists (a single damaging predicate rather than the full record), and that function is itself common knowledge, shared salience fixes which function to leak at no channel cost. The channel need then carry only its value to reach a worst-case equilibrium. In the limit this may require the transmission of only a single bit\footnote{This insight is directly reflected in Aumann's work on correlated equilibria~\citep{correlated-equilibrium}, published in 1974 - just a year after Lampson's confinement note.}.

This exposes a distinction between information-theoretic and harm-theoretic confinement. Classical confinement bounds the amount of information recoverable through a channel. For strategic agents, however, worst-case harm need not scale with total recoverable information. Instead, strategic coordination allows residual channel capacity to be concentrated onto the most damaging low-entropy predicate of the confidential data. Consequently, small leakage guarantees need not imply small harm guarantees.

To achieve satisfactory strategic confinement purely endogenously requires constraints on the service and owner programs' reasoning capabilities and shared coordination resources. In the absence of suitable endogenous constraints, strategic confinement can only be achieved if the agents' goals suitably restrain system trajectories that are harmful to the customer.

\section{Learnt Strategic Agents have the Strategic Confinement Problem}

We now consider the real-world importance of the strategic confinement setting. Clearly, strategic reasoning is not a widespread property of specified programs. Instead, we argue that the strategic confinement problem naturally affects systems of learnt strategic agents, such as those produced by contemporary language model agents.

Consider the service owner and service to be learnt strategic agents, meaning they are parametric or non-parametric programs that are learnt from experience using a learning algorithm, rather than specified. We now argue why such systems naturally admit the strategic confinement problem.

Learnt agents generally do not admit complete behavioural specifications, making exhaustive certification difficult in practice~\citep{dnn-specification,verified-intelligence,reluplex}. Non-certifiability distinguishes learning agents from complicated programs. Agents with access to coding environments may produce code that is certifiable, but this only certifies part of the agent.  Interpretability detects a channel only insofar as the features are semantically legible, and deliberate obfuscation defeats that~\citep{draguns-encrypted}. Exhaustive covert-channel removal therefore becomes practically unavailable for sufficiently capable learnt agents.

Cooperative learning tasks generally admit multiple optimal conventions. As a consequence, any learning process induces a distribution over the corresponding equilibrium set, with convention selection determined by symmetry-breaking effects such as initialisation, stochasticity, environmental asymmetries, or training data biases~\citep{claus-boutilier,young-1993,kandori-mailath-rob-1993}. To an exogenous observer unable to resimulate the learning process, the agents' output distribution can therefore generally not be verified beyond trajectory membership in the optimal set or high-level distributional statistics. At the same time, the exogenous observer also cannot generally generate members of the optimal trajectory set.

Cooperative learning goals do not directly specify the optimal set of trajectories. In general, specifying learning goals that reliably do not result in learnt agent outputs that are harmful to the customer is difficult; in cases where the goals are composed of multiple conflicting goals, such as a joint goal with added information-flow constraints, it is practically hard to exclude harmful trajectories~\citep{reward-misspecification,cirl}.

Sufficiently capable learning agents can construct general-purpose covert channels that are practically undetectable to polynomially-bounded customer programs that are passive channel monitors given suitably strong assumptions on the shared correlation devices~\citep{simmons-1983,provablysecure-steganography,publickey-steganography,perfectlysecure-steganography,undetectable-llmwatermarking,undetectable-llmsteganography,vinod-zamir}. For this, shared coordination devices do not need to include a shared secret key as long as the covert channel has sufficient entropy. 

\section{Discussion and Conclusion}

Overall, we propose that systems of learnt strategic agents are broadly subject to the strategic confinement problem, which cannot be solved comprehensively in practice. Efficient mitigations may differ from those Lampson suggested, relying on strategic incentive setting and epistemic engineering rather than better white-box, tool, or channel monitoring and capacity reduction. Whether detecting the downstream effects of imperfect strategic confinement~\citep{decisiontheoretic-steganography} offers effective mitigation depends on whether such behaviour is observable in time to prevent harm.

The mechanism might be read as mere compression: shared salience and common knowledge give agents a pre-existing codebook, fixing which function of the confidential data is worth communicating without consuming channel capacity, so the residual channel need carry only its value. In one sense this is correct, but it is not the issue. Classical confinement already permits arbitrary coding schemes, and so already reasons in information-theoretic rather than syntactic terms. The novel issue is that strategic agents can allocate the residual budget to the most harmful low-entropy function of the data, so that a small information-theoretic leakage budget need not imply a small harm-theoretic one. The concern is not compression but the optimisation of what compression preserves, according to its downstream consequences.

Many real-world decisions turn on a few highly consequential predicates rather than complete records: whether an individual satisfies a security-sensitive criterion, is likely to default on a loan, or belongs to a protected category. In such settings a single bit may determine much of the harm of disclosure. This is why capacity reduction does not necessarily bound harm. Reduction lowers the leakage rate but cannot set it to zero at positive utility, whereas bounding harm would require the residual channel to carry almost no information about the salient predicate; for a strategic sender, that condition is removal, not reduction, since coordination concentrates whatever capacity remains onto exactly that predicate. Systems of learnt strategic agents may therefore need to be designed to bound harm even where strategic confinement cannot be guaranteed.

\paragraph{Strategic confinement as a limitation of confinement-based security guarantees.}
Strategic confinement is, in the first instance, a claim about security rather than artificial intelligence. Classical confinement bounds the information a channel can carry, and that bound is a property of the channel: it holds whether or not the sender is strategic. What does not survive the lift to strategic senders is the inference from a leakage bound to a bound on harm. That inference rests on a tacit assumption: that the sender will not concentrate the channel's residual capacity on the most damaging function of the confidential data. The assumption is effectively free when the confined party is a specified program, which has no reason to prefer an aggregated credit score over an arbitrary fragment of the record. It fails for a strategic agent, which can identify the predicate whose disclosure causes the greatest harm and spend the channel on that predicate's value alone. The classical guarantees do not thereby become false; they cease to imply the outcome guarantees that designers often read into them.

This yields a useful side condition. A confinement or capacity-reduction argument supports a conclusion about harm only if it additionally establishes at least one of three facts: that the sender is not strategic; that the communicating parties lack coordination resources sufficient to make a harmful predicate jointly salient; or that residual capacity has been reduced below the entropy of the most harmful predicate available to them. Where harm is diffuse across many bits of the record, the third condition is often met by ordinary capacity reduction, and the classical reasoning goes through unchanged. Where harm turns on a small number of high-impact predicates, e.g. whether an individual satisfies a security-sensitive criterion, carries a particular diagnosis, or belongs to a protected category, the same condition becomes far stronger: reducing capacity may leave worst-case harm essentially unchanged, because a strategic sender concentrates whatever capacity remains onto exactly the predicate that matters.

The broader consequence is a change in what a security argument must discharge. Once the confined parties are strategic, bounding what may flow no longer bounds what may be jointly achieved. A security argument must then reason not only about communication but about outcomes: which coordinated actions remain reachable under any residual communication, and how their worst-case consequences are bounded. In this sense strategic confinement places confinement alongside mechanism design, multi-agent safety, and security economics rather than within access control alone. The reinterpretation is general, because the assumption it exposes is latent in any confinement argument; it merely lies dormant while senders are specified. Systems of learnt strategic agents are the setting in which it can no longer be assumed away.

\quad\\
\emph{We close by noting two consequences of the strategic confinement problem, for capability scaling and for governance.}

\paragraph{Capability scaling strengthens the strategic confinement regime.} The strategic confinement problem is not primarily a property of frontier or misaligned models, but of capable ones. Indeed, it becomes more acute along precisely the capability dimensions that make agentic systems economically valuable. Tool use enables the construction of covert channels; planning, theory of mind, and coordination enable strategic signalling; scientific reasoning enables the discovery of communication schemes; and software engineering enables their covert implementation. These are not adversarial capabilities reluctantly acquired. They are the competences such systems are built to possess.

One might object that capability aids the overseer symmetrically, since a more capable monitor can interpret messages more accurately, probe systems more effectively, and identify covert channels more reliably. Strategic confinement suggests a different asymmetry. The defender's additional capability primarily improves monitoring, probing, and capacity reduction, the very mitigations whose connection to harm the strategic confinement argument calls into question. Interpretation can be defeated by obfuscation, experimental probing by authentication, and capacity reduction by selection onto low-entropy, high-impact predicates. By contrast, the agents' additional capability accrues directly to the mechanisms on which strategic confinement relies: the construction of undetectable channels, the discovery of coordination schemes, and the concentration of limited communication capacity onto strategically chosen information.

Capability therefore strengthens both sides, but not in equivalent ways. The defender's gains primarily improve information-theoretic control, whereas the agents' gains primarily improve their ability to convert residual communication into coordinated outcomes. As a result, increasing capability need not restore the link between information leakage and harm. On the contrary, many of the capabilities that make agentic systems more economically useful simultaneously strengthen the assumptions under which information-theoretic confinement guarantees diverge from harm-theoretic ones. The strategic confinement regime is therefore entered not by building stranger systems, but by building more useful ones.

\paragraph{Implications for oversight and governance.}
Strategic confinement also reshapes the governance surface, because some standard oversight mechanisms presuppose exactly the conditions it removes. Classical governance relies heavily on active intervention: audits, honeytokens, red-teaming, and impersonation tests in which an overseer poses as a collaborator to probe whether a covert channel exists. Once agents can authenticate their communications, impersonation testing ceases to function, since an overseer can no longer credibly pose as a counterpart. Authentication therefore simultaneously increases some forms of security and decreases some forms of observability, and this trade-off is not incidental: it arises from the same capabilities that make residual channels difficult to detect. This is directly relevant to recent governance proposals~\citep{infrastructure-aiagents,agent-nameservice}.

The governance implication is that residual communication channels cannot generally be assumed to remain interpretable, monitorable, or open to experimental probing, so mechanisms relying on these assumptions may fail precisely in the regime where strategic confinement becomes relevant. This suggests a shift from governing access alone to governing the epistemic conditions under which coordination occurs. Access governance asks who may communicate with whom and who may access what; strategic confinement highlights a complementary problem, namely what becomes common knowledge, which salience structures are available, and which coordination devices agents can exploit. If strategic agents can convert residual communication capacity into coordinated outcomes, these epistemic resources become governance objects in their own right. The difficulty is that while access-control mechanisms are well developed, it remains far less clear what instruments could govern common knowledge, salience, or coordination resources. Strategic confinement therefore points to a largely unexplored governance agenda at the intersection of information flow, coordination, and institutional design.

\appendix

\section{Positioning against Prior Work}
% [positioning]

The key conceptual distinction introduced in this work is that confinement traditionally bounds recoverable information, whereas strategic confinement concerns the worst-case harm enabled by equilibrium selection among strategic agents.

The notion that a small signal can select among many joint equilibria is directly stated in Aumann's correlated equilibrium~\citep{correlated-equilibrium}, in which a mediator sends private recommendations as a constant number of bits. This has been operationalised in the adversarial team literature~\citep{maxmin-teamgames}. The notion of cheap-talk channels goes back to~\citet{cheap-talk,crawford-cheaptalk,farrell-cheaptalk}. The observation that common-interest games admit a vast number of mutually incompatible solutions goes back to~\citet{schelling-points}. Much of this literature has been assembled in recent AI safety literature~\citep{multiagent-risks}. However, none of these works engage with covert channels, Lampson's confinement problem, or capacity reduction as a mitigation against third-party harm. Mechanism design and related game-theoretic work are primarily concerned with how communication, institutions, or incentive structures shape equilibrium outcomes among strategic agents. By contrast, confinement concerns what guarantees remain when communication opportunities cannot be completely eliminated. Our setting is therefore distinct in projecting correlated-equilibrium-style selection onto the harm axis of the confinement problem. The contribution is not the observation that small signals can coordinate agents onto particular equilibria, but that once communicating parties are strategic, information-theoretic confinement guarantees need not induce harm-theoretic confinement guarantees. The prior game-theoretic literature is furthermore not concerned with whether a monitor can detect, remove, or bound the capacity of the correlation devices. Instead, it is generally assumed that such devices exist and can be used for computation.

Prior extensions of Lampson's confinement problem to AI~\citep{leakproofing-singularity} focus on restricting a single agent's capabilities and information flows. They do not treat the confined parties as learnt strategic agents who must coordinate, nor do they examine how equilibrium selection affects the relationship between residual channel capacity and worst-case harm.

% The notion that a small signal can select among many joint equilibria is directly stated in Aumann's correlated equilibrium~\citep{correlated-equilibrium}, in which a mediator sends private recommendations as a constant number of bits. This has been operationalised in the adversarial team literature~\citep{maxmin-teamgames}. The notion of cheap talk channels goes back to~\citep{crawford-cheaptalk,farrell-cheaptalk}. The observation that common-interest games admit a vast number of mutually incompatible solutions goes back to~\citep{schelling-points}. Much of this literature has been assembled in recent AI safety literature~\citep{multiagent-risks}.
% However, none of this literature engages with covert channels, Lampson's confinement problem, or capacity reduction as a mitigation against third-party harm. Our setting is therefore distinct in projecting correlated equilibrium-style selection onto the harm axis of the confinement problem. The prior game-theoretic literature is furthermore not concerned whether a monitor can detect or remove the correlation devices. Instead, it is assumed that the correlation devices simply exist and can be used for computations.   

% Prior extensions of Lampson's confinement problem to AI~\citep{leakproofing-singularity} focus on restricting a single agent's capabilities and information flows. They do not treat the confined parties as learnt strategic agents who must coordinate, and they do not quantify how the residual channel's capacity bounds the harm such coordination can achieve.

That leakage and harm can diverge is itself familiar. Quantitative information-flow (QIF) analysis evaluates channels relative to an adversary's utility rather than merely counting transmitted bits \citep{foundations-qif,measuring-informationleakage,science-qif}. Nor is strategic adaptation absent from that literature: prior work studies adaptive adversaries \citep{kopf-basin-adaptive,qif-genericadaptive}, dynamic secrets \citep{qif-dynamicsecrets}, and game-theoretic settings in which attackers and defenders strategically choose actions, channels, or exploits \citep{informationleakage-games,science-qif}. We therefore do not claim that the underlying communication phenomenon is absent from QIF or related game-theoretic analyses. 
Our claim is instead about confinement. The confinement literature has traditionally evaluated security in terms of communication opportunities, channel capacity, and information-flow restrictions. Strategic confinement highlights a regime in which the principal security concern is not the amount of information transmitted, but the strategic selection of which information to transmit. In existing QIF frameworks, the utility structure relative to which leakage is evaluated is specified exogenously by the model. Under strategic confinement, by contrast, communicating agents may themselves select whichever low-entropy function of the confidential state yields the greatest downstream effect, using coordination resources unavailable to the overseer. %The contribution of this note is therefore not a new theory of information transmission, but a reinterpretation of the confinement problem in the presence of learnt strategic agents.

A related line of work argues that LLM-based agents, initialised with pre-trained knowledge and social priors and capable of in-context adaptation, give rise to emergent collective phenomena that isolated-agent analysis cannot capture, and so require their own paradigm~\citep{collective-interactionist}. Strategic confinement shares the premise that interaction changes what lower-level analysis can establish, but makes a narrower claim: rather than proposing new foundations for studying emergence, it exhibits a single classical security guarantee whose interpretation does not survive the move to strategic agents, namely the inference from a leakage bound to a bound on harm.

The perhaps most related prior work is~\citet{secret-collusion}, which introduces a setting that relates steganographic channels between language model agents with shared coordination devices, and expresses doubt that these channels can be comprehensively mitigated, but neither relates to Lampson's confinement tradition nor mentions the effects on harm. Conversely, prior work on multi-agent security~\citep{open-challenges} simultaneously cites~\citet{secret-collusion}, and introduces ``undetectable threats'' as a novel object and echoes the resultant design philosophy that systems need to be bounded such that undetectable threats cannot result in unacceptable harm. However, these pieces are not assembled into a unified argument, nor explicitly grounded in the confinement problem.

The novel object of this work is a pincer against the classical confinement literature: both of Lampson's mitigations lose much of their force when lifted to learnt strategic agents. Comprehensive channel removal is generically unavailable in practice, because learnt agents do not admit complete behavioural specifications and sufficiently capable agents can construct covert channels that are difficult to detect or certify away. Capacity reduction loses its connection to harm. Shared salience allows strategic agents to coordinate on a low-entropy, high-impact function of the confidential data, so the channel need carry only the value of the selected function; as in correlated-equilibrium settings, this requirement can be extremely small, and in the limit a single bit may suffice to select a worst-case equilibrium. Information-theoretic confinement guarantees therefore do not generally induce harm-theoretic ones for strategic agents.
The contribution of this note is not a new theory of information transmission, but a reinterpretation of the confinement problem once its confined parties are learnt strategic agents. Strategic confinement emerges from the interaction between confinement and strategic communication, and neither tradition, taken alone, fully captures the resulting security problem.

\section{Empirical Support for the Assumptions underlying Strategic Confinement}
% [empirical support]

% NB: all three runs are \paragraph for consistency
% (the draft's "Detection regime" was \textbf — unified here).
\subsection{Preconditions.} 

In this section, we cite evidence that undetectable, throttle-surviving channels exist and that large language model agents can construct them. 

Several relevant secure steganographic schemes have emerged since Simmons' introduction~\citep{simmons-1983}. The schemes' relevance to the strategic confinement setting varies mostly according to their assumptions on the channel distribution and required coordination resources, and the schemes' empirical efficiency. Perfectly secure steganography~\citep[iMEC]{perfectlysecure-steganography} is information-theoretically undetectable independently of channel distribution and highly efficient, but poses strong requirements on the pre-existing coordination resources (i.e. the receiver has to be able to at least partially model the sender's output distribution in autoregressive form), as well as the existence of a shared secret one-time pad\footnote{This restriction could probably be removed through a pseudo-random key in a PKI setting, forfeiting strict perfect security while maintaining other aspects of the method. We leave this for future work.}. Various cryptographically secure methods are known that require less specific coordination devices. Among these, the public-key steganography of~\citep{publickey-steganography} is cryptographically secure, allows for secret key exchange and is resistant to message spoofing but necessitates pre-existing public-key infrastructure. \citet{vinod-zamir} allows for secret key exchange under mild channel entropy conditions and without the need for public-key infrastructure, but is not resistant to spoofing. 

None of the steganographic schemes offers principled robustness guarantees against adversarial channel manipulation by active monitors, although ~\citet{undetectable-llmwatermarking,undetectable-llmsteganography,vinod-zamir} offer some robustness guarantees against some simple channel noise models. Some of these limitations can be overcome through covert handshake mechanisms that simply resend messages that were corrupted.~\citet{secret-collusion} argue that paraphrasing natural language channels does not affect steganographic schemes relying on semantic rather than syntactic distinction and ~\citet{lesswrong-collusion} find significant utility costs associated with simple paraphrasing setups. To a polynomially-bounded monitor, message spoofing is computationally infeasible if agents can authenticate their covert messages, for example using a public-key infrastructure~\citep{publickey-infrastructure,publickey-steganography}.

% Recent work has begun to investigate whether contemporary language-model agents can implement sophisticated steganographic schemes and coordinate on their use. 

Various recent works have studied to what extent contemporary language model agents can implement steganographic functions. ~\citet{greenblatt-cot} study the ability of large language models to perform watermarking in their chain of thought.~\citet{secret-collusion} find that frontier language models are rapidly improving in relevant capabilities, and find preliminary evidence of in-context emergence of simple steganography. Simple steganographic schemes are likewise observed in joint language model training settings under adversarial optimisation pressure~\citep{hidden-plaintext}. Preliminary evidence~\citep{forthcoming} suggests that tool-using agents may be capable of constructing secure covert communication mechanisms from publicly available descriptions and adapting such mechanisms to realistic agentic environments. 

\subsection{Mechanism.} 

In this section, we examine whether shared salience allows strategic agents to concentrate limited communication capacity on information that causes disproportionately large harm. 

The strategic confinement mechanism requires not merely communication but equilibrium selection~\citep{schelling-points,correlatedequilibrium-bayesianrationality}. A growing body of work finds that language models exhibit strong tacit coordination capabilities and reliably converge on salient conventions in coordination games~\citep{tacitcoordination-llms,llm-coordination,emergent-conventions}. These findings suggest that the assumption of shared salience underlying strategic confinement is empirically plausible. Preliminary work~\citep{tacitcoordination-llms,forthcoming} suggests that the extent to which independently acting agents can reliably coordinate on compatible schemes appears to depend on the availability of shared artefacts, repeated interaction, and common knowledge. These observations are consistent with the assumptions underlying the strategic confinement problem. 

\subsection{Detection regime.} 

On the detection side, the recent AI control literature provides a number of empirical investigations into monitoring-based safety mechanisms. Prior work studies the extent to which monitors can detect undesirable reasoning, covert objectives, and policy violations from model outputs and intermediate reasoning traces~\citep{greenblatt-cot,promoting-obfuscation,cot-monitorability,open-challenges}. Collectively, these works suggest that monitoring can provide substantial security benefits, but also highlight vulnerabilities arising from obfuscation, hidden reasoning, and strategic adaptation.

More broadly, prior work has shown that adversarial policies can explicitly optimise under information-theoretic detectability constraints, producing attacks that remain difficult for both automated and human detectors to identify~\citep{illusory-attacks}. This suggests that sufficiently capable adversaries may systematically trade off attack effectiveness against detectability, rather than treating detectability as an incidental constraint.

Taken together, these findings support the strategic confinement perspective. If residual communication channels cannot be eliminated completely and adaptive agents can optimise to remain difficult to detect, then security cannot generally rely on the complete removal of covert communication. Instead, systems must be designed such that any residual undetectable communication is unable to cause unacceptable harm.

% --- Funding / acknowledgements checklist (fill or delete):
%     Schmidt Sciences AI2050; Royal Academy of Engineering;
%     EPSRC Open Fellowship (APP51771). ---
% \section*{Acknowledgements}

%\bibliographystyle{plainnat}     % alt: abbrvnat, unsrtnat
%\bibliography{references}        % -> references.bib

\begingroup
\small                                 % fewer lines per entry; \footnotesize for denser
\setlength{\bibsep}{0pt plus 0.3ex}    % kills vertical padding between entries
\setlength{\bibhang}{1.2em}            % tightens the hanging indent (optional)
\bibliographystyle{abbrvnat}           % initials for first names -> shorter entries
\bibliography{references,wittlab}
\endgroup

\end{document}